\documentstyle[aps,prl,twocolumn]{revtex}
\input{rotate.tex}
\input{epsf.tex}

\begin{document}  


\draft 
\title{Ballistic Transport in Superconducting Weak Links in a Microwave Field} 
\author{U. Gunsenheimer$^a$, and A.D. Zaikin$^{b,c}$}  
\address{$^a$ Fakult\"at f\"ur Physik und Astronomie, Lehrstuhl f\"ur 
Theoretische Festk\"orperphysik,\\
 Ruhr--Universiti\"at Bochum, 44780 Bochum, Germany}  
\address{$^b$ Institut f\"ur Theoretische Festk\"orperphysik, Universit\"at 
Karlsruhe, 76131 Karlsruhe, Germany}  
\address{$^c$ I.E. Tamm Department of Theoretical Physics, P.N. Lebedev 
Physics Institute, \\
Leninsky Prospect 53, Moscow 117924, Russia} 

\maketitle  

\abstract{Nonequilibrium effects and their impact on a charge
transport in superconducting ballistic weak links biased by an ac voltage 
are investigated within the framework of the Keldysh technique. 
We demonstrate that the microwave field destroys the phase coherence 
during the multiple Andreev reflection cycle and leads 
to the effective $cooling$ of subgap quasiparticles accelerated
due to multiple Andreev reflection. For small bias voltages 
this effect results in a strong supression of both the excess current 
and the conductance of the weak link. In the opposite
limit of large bias voltages the excess current remains unaffected.
We also demonstrate that a simple Boltzmann kinetic approach
becomes inadequate if an ac voltage bias is applied to the weak link.}  



Multiple Andreev reflection leads to excitation of quasiparticles in voltage 
biased superconducting weak links \cite{OTBK}. As a result the quasiparticle 
distribution function for such systems is driven out of equilibrium already at 
small voltages. In the case of a time independent voltage bias charge transport 
in ballistic superconducting weak links in the presence of such nonequilibrium 
effects has been studied by Octavio et al. \cite{OTBK} within the framework of 
a simple classical Boltzmann kinetic equation. This so called OTBK model was 
then widely used in a large number of experimental as well as theoretical works. 

Although the OTBK model provides a transparent physical picture of multiple 
Andreev reflection and dissipative charge transport in superconducting weak 
links it remained unclear if (and/or under which conditions) this model is 
sufficient to describe quantum nonequilibrium effects in such systems. More 
recently a rigorous theory of charge transport in ballistic 
superconductor--normal metal--superconductor (SNS) structures in the presence 
of a $constant$ voltage bias has been developed by means of the Keldysh 
technique \cite{GuZai94,AvBard96}. In the absence of inelastic relaxation the 
result for a time--averaged dissipative current across the system obtained in 
\cite{GuZai94,AvBard96} $exactly$ coincides with that derived from the OTBK 
model thus providing a formal justification for the OTBK results \cite{OTBK}.

How far can one go applying the OTBK model to various nonequilibrium
effects in superconducting weak links? Is the agreement between the
results \cite{OTBK} and \cite{GuZai94,AvBard96} specific for $ballistic$
weak links biased by a $dc$ voltage or further
generalization of the OTBK model (see e.g. \cite{ZimKeck96}) is possible?

In this Letter we will study nonequilibrium effects in superconducting 
microconstrictions in the presence of a time-dependent voltage within the
framework of the Keldysh technique. We shall determine the distribution 
function and obtain the current voltage characteristics (CVC) of the weak 
links. We will demonstrate that photon absorbtion and emission processes  
in the weak link (in combination with multiple Andreev reflection)
may have a strong impact on the transport properties of the system
leading to effective heating or cooling of subgap quasiparticles and 
to a strong suppression of the current in the limit of small bias voltages. 
The latter effect may be of interest for applications of SNS structures
as microwave detectors. We will furthermore argue that such nonequilibrium 
effects cannot be adequately described within the (generalized) OTBK model 
\cite{OTBK}.  


{\it The model.} 
Nonequilibrium effects in inhomogeneous superconductors are conveniently 
described by means of quasiclassical Greens functions in 
Keldysh--Nambu space \cite{LarOv76+}, 
 \begin{equation} \label{G}
    \hbar \vec{v}_{F} \nabla {\check G}
  + {\hat\sigma}_{3} \hbar \frac{\partial}{\partial t}{\check G}
  + \hbar \frac{\partial}{\partial t'}{\check G}{\hat\sigma}_{3}
  + \left[ {\check K}, {\check G}\right] = {\check 0 },  
\label{Eileq}
 \end{equation}
where ${\check G}$ is a $2\times 2$ matrix in the Keldysh space consisting
of retarded, advanced and Keldysh Green functions $\hat G^{R,A}$ and $\hat
G^K$. The latter are in turn $2\times 2$ matrices in Nambu space. 
$\check G$ obeys the normalization condition $\check G \circ \check 
G = \check 1 \delta(t-t')$ where the "$\circ$" product indicates integration 
over the internal time variable. The same integration is also implied in the 
commutator $[\check K, \check G]$ where ${\check K}(\vec{r},t,t') = 
{\hat H} (\vec{r},t)\check 1 \delta(t-t') + i{\check\Sigma}(\vec{r},t,t')$
with $\hat H(\vec{r},t) = i[U(\vec{r},t){\hat 1} - \hat \Delta (\vec{r},t)]$. 
Here, $U$ is the scalar potential, and $\Delta$ is the off--diagonal pair 
potential. 

We consider a standard model of a short superconducting microconstriction 
(or bridge): two superconducting bulks are connected directly or via a small 
piece of a normal metal of a typical size much smaller than the superconducting 
coherence length $\xi_0$ (see e.g. \cite{GuZai94}). We will assume that the 
external voltage $V(t)$ applied to the system drops at the junction, and the
electric field does not penetrate into superconducting electrodes. The 
superconducting phase difference across the junction is then defined in 
a standard way $\dot \varphi(t)=2eV(t)/\hbar$. 


{\it Nonequilibrium distribution function}.
We follow Ref. \cite{GuZai94} and construct the Green functions
by solving Eqs. (\ref{Eileq}) in each superconductor and matching 
these solutions continuously at the contact interface. 
The quantum kinetic properties of the system are 
completely described by the Keldysh Green function $\hat G^K$ which can be 
expressed as $\hat G^K = \hat G^R \circ\hat h - \hat h\circ\hat G^A$ with  
$\hat h = 1 - 2f - 2\tilde f\;\hat\sigma_3$ \cite{LarOv76+}. 
With $\hat G^K$, 
$\hat G^R$ and $\hat G^A$ known we have constructed $f$ and $\tilde f$. 
The Fourier transforms of the latter with respect to time difference 
represent the ``longitudinal'' and ``transversal'' 
components of the distribution function describing respectively energy 
and charge modes \cite{Schmid}, $\hat \sigma_3$ is the Pauli matrix. In 
equilibrium $f$ is equal to the Fermi distribution function $f_0$ and 
$\tilde f =0$. A striking feature
of {\it ballistic} SNS bridges is that inside the N-metal the function 
$\tilde f $ is equal to zero even if $f$ is driven out of equilibrium
due to multiple Andreev reflection in the presence of an externally applied 
voltage. The key reason for this result is that the electric field does not 
penetrate into the ballistic N-metal (the voltage drops only across NS
interfaces \cite{GuZai94,Zaikin83}) and therefore does not cause charge 
imbalance responsible for a nonzero $\tilde f$ component
of the distribution function \cite{VolKlap92}. As the latter component cannot 
be described by means of a classical Boltzmann equation we conclude that 
the condition $\tilde f=0$ is an important prerequisite for the validity of 
the OTBK model. 

In the limit of a constant voltage $V(t)=\bar V$ applied to the system the 
distribution function $f$ does not depend on time and is given by $f \equiv 
f^{(0)}_{\pm}(E) = \sum_{n=0}^{\infty} f_{\pm,n}^{(0)}(E)$ with 
 \begin{eqnarray} \label{fn0}
  f_{\pm,n}^{(0)}(E) &=& \prod_{l=0}^{n-1}{\cal A}(E\!\mp\!le\bar V)
   \left[1\!-\!{\cal A}(E\!\mp\!ne\bar V)\right]f_0(E\!\mp\!ne\bar V).
\end{eqnarray}
Here ${\cal A}(E) \equiv |\gamma^R(E)|^2$ is the Andreev reflection 
probability where (neglecting inelastic scattering) $\gamma^R(E) = 
(E-\sqrt{E^2-\Delta^2})/\Delta$. ``$+$'' and ``$-$'' label quasiparticles 
with momentum in and opposite to the direction of the current flow. Due to 
multiple Andreev reflection in the presence of electric field ``$+$'' 
quasiparticles are accelerated and the distribution function (\ref{fn0}) 
for energies within the gap increases, i.e. the effect of {\it heating} 
of such quasiparticles takes place. It is accompanied by a symmetric effect 
of {\it cooling} of ``$-$'' quasiparticles which distribution function for 
subgap energies is suppressed.  

Note that in the absence of inelastic relaxation 
(which causes deviations of ${\cal A}(E)$ from its BCS value) 
the expression (\ref{fn0}) {\it exactly} coincides with that 
obtained by OTBK \cite{OTBK}.
This coincidence is by no means surprising: electrons and holes suffer 
no scattering in a clean N-metal and obviously can be described by 
the classical Boltzmann equation. Taking into account Andreev reflection
at NS interfaces by imposing proper boundary conditions \cite{OTBK} 
one arrives at the results equivalent to those obtained within
the general quantum kinetic analysis \cite{GuZai94,AvBard96}.

The situation changes in the presence of an additional microwave field 
$V(t)=\bar V+\tilde V\cos \omega t$. In this case electrons and
holes moving in the N-metal can absorb and emit photons. Obviously
such processes cannot be correctly described by the Boltzmann equation 
which does not contain information about off--diagonal elements of
the density matrix. Here we elaborate the quantum kinetic
analysis (\ref{Eileq}) and evaluate the distribution function $f$ 
which now becomes time dependent. In a Fourier representation 
$f_{\pm}(E,t) = \sum_{\kappa=-\infty}^{\infty} 
f_{\pm,\kappa}(E)\;\exp(\pm i\kappa\omega t)$ we find 
 \begin{eqnarray} \label{f-kappa}
  & & f_{\pm,\kappa}(E) = [1-{\cal A}(E)]\,f_0(E)\,\delta_{\kappa,0} \\ 
  &+& \sum_{k=-\infty}^{\infty} \, \sum_{\kappa'=-\infty}^{\infty} \, 
    m_{\kappa,\kappa'}(\pm,k,E)\; 
    f_{\pm,\kappa'}(E\!\mp\!e\bar{V}\!\mp\!k\hbar\omega). \nonumber 
\end{eqnarray}
Here the terms containing the coefficients  
 \begin{equation} \label{mkk}
 m_{\kappa,\kappa'} = (\pm i)^{\kappa-\kappa'}\,J_k\,J_{k-(\kappa-\kappa')}
                             \gamma^R(E)\,[\gamma^R(E\mp\kappa\hbar\omega)]^*
\end{equation}
take care about all possible photon absorption and emission processes
(the probability amplitude for $k$--photon processes is given
by the Bessel function $J_k \equiv J_k(e\tilde V/\hbar \omega)$).

Eqs. (\ref{f-kappa}) and (\ref{mkk}) selfconsistently determine the 
quasiparticle distribution function for a weak link in the presence of an 
ac voltage. This is one of the main results of the present paper. It is 
important to point out that terms with $\kappa \neq 0$ contain information 
about the phase shift of electron and hole amplitudes due to photon absorbtion
and emission and cannot be recovered from the Boltzmann equation analysis 
\cite{ZimKeck96}.

We have solved Eqs. (\ref{f-kappa})--(\ref{mkk}) selfconsistently for 
various values of $\omega$, $\bar V$, $\tilde V$ and $T$. The results 
for the time independent part of the distribution function $f_{\pm,\kappa=0}$ 
are shown in Fig.~1. 
\begin{figure}[t]
\epsfbox{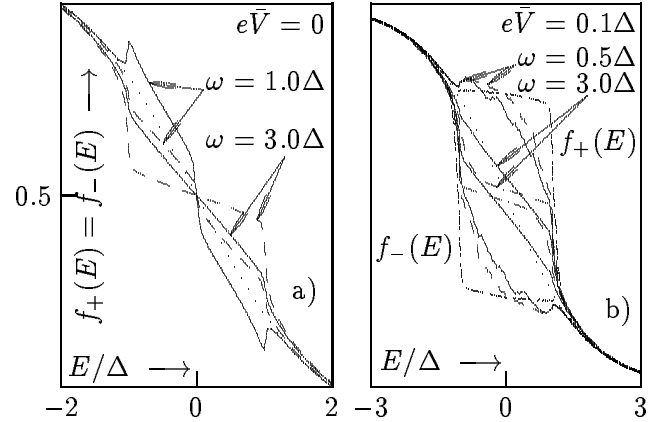}
\caption{Distribution function $f_{\pm}(E)$ at $T=0.9\,T_C$ for 
a) $e\bar V = 0$, b) $e\bar V = 0.1\,\Delta$ and different $\omega$. 
Solid curves: finite microwave field $e\tilde V/\hbar\omega=1$, 
dashed curves: results according to Ref. [4], dashed--dotted curves: 
zero microwave field, dotted curves: Fermi distribution $f_0$.} 
\end{figure}
Our numerical analysis captures all essential features which can be
summarized as follows.

In the case $\bar V=0$ (Fig.~1a)) the microwave field excites out of the Fermi 
surface. At frequencies $\hbar\omega<2\Delta$ and small microwave amplitudes
only single photon processes are important and we find an increase of $f_+(E)=
f_-(E)=f(E)$ at energies $-\Delta < E < 0$ and a decrease at $0<E<+\Delta$. On 
top of that $f(E)$ exhibits a peak at $|E|{\scriptstyle <\atop\sim}\Delta$. This 
behavior is a consequence of inelastic quasiparticle scattering into (and out of) 
the bound states in the short constriction, $E_{\pm}=\pm \Delta\cos(\varphi/2)$, 
due to photon absorbtion. The small alternating phase difference $\varphi(t)$ 
leads on the one hand to a certain smearing of the bound state level around 
$E_{\pm} =\pm\Delta$. On the other hand photon absorbtion of quasiparticles in 
the $E_+$ state yields an emptying of the latter. Symmetrically, quasiparticles 
with energies $E<-\Delta$ absorb a photon and fill up the $E_-$ state. It is 
interesting that due to this resonance effect the subgap distribution function
may be tilt to the right which corresponds to an effective $cooling$. For 
frequencies $\hbar\omega > 2\Delta$ the sign of the change in the distribution 
function, $\Delta f$, is altered and we find an effective $heating$. In this case 
quasiparticles with $E\leq E_-<0$ may be scattered into the energy range $E>0$. 
In addition we note here (without showing a figure) that also for $\hbar\omega<
2\Delta$ the sign of $\Delta f$ may be changed if the microwave amplitude $\tilde 
V$ is larger so that multi photon processes become important.  

For nonzero $\bar V$ we observe a nontrivial combination of two effects:
acceleration of quasiparticles due to multiple Andreev reflection (cf. 
Eq. (\ref{fn0})) and their excitation by a microwave field. The
former effect depends on the momentum direction whereas the latter is
sensitive to the sign of the energy $E$. As a result the structure 
of the distribution function for subgap energies turns out to be quite 
complicated. This function is shown in Fig.~1b) (solid lines) for two
different frequencies of the microwave field. For small $\omega$ (see 
curve with $\omega=0.5\Delta$) we observe a relative $cooling$ of the 
``$+$'' branch with respect to the case $\tilde V=0$ (the latter curve 
is depicted by dashed-dotted lines). The physical reason for this effect 
is clear: photon absorbtion and emission processes break the multiple 
Andreev reflection cycle thus preventing from further acceleration of 
quasiparticles. But still the acceleration effect dominates for such 
$\omega$ and the ``$+$'' branch remains overpopulated. On the other 
hand, for larger values of $\omega$ (see curve with $\omega=3\Delta$) 
photon absorbtion dominates over multiple Andreev reflection (for 
$\omega > 2\Delta$ quasiparticles are exited outside the subgap energy 
interval already due to one photon processes) and the distribution 
function becomes closer to that for $\bar V =0$. Still some dependence 
of $f$ on the momentum direction remains.

The above physical picture is correct as long as $e\bar V \ll \Delta$,
i.e. subgap quasiparticles suffer many Andreev reflections. For
large voltages $e\bar V >2\Delta$ only one Andreev reflection is possible
and the microwave field effect less important. For a wide interval of
voltages we observe additional subharmonic gap structures with the period
$\hbar \omega$ (Fig.~1b) or satellite structures with $\delta E=\pm\hbar\omega$ 
in the vicinity of the usual subharmonic gap structures (figures not shown). 

As we already pointed out the problem has been recently investigated 
by Zimmerman and Keck \cite{ZimKeck96} within the classical Boltzmann
kinetic equation approach extending the OTBK analysis \cite{OTBK}
to the ac voltage situation. This approach essentially deals
only with diagonal elements of the density matrix and contains no information
about off-diagonal phase sensitive terms. Our results reduce to those
of Ref. \cite{ZimKeck96} provided these terms ($m_{\kappa,\kappa'}$ with 
$\kappa\neq 0$ in Eq. (\ref{f-kappa})) are neglected. At low voltages these
terms are significant, and the results \cite{ZimKeck96} (dashed lines in Fig.~1)
considerably (and for $\hbar\omega<2\Delta$ even $qualitatively$) deviate from 
those obtained here. On the other hand for larger $e\bar V\sim\Delta$ only a 
few Andreev reflections are possible, off--diagonal elements play a minor role 
and the agreement between our results and those of Ref. \cite{ZimKeck96}
is better. 


{\it Current-voltage characteristics.} The current across the weak link is 
determined by the expression for the Keldysh Green function $\hat G^K$ in 
a standard way (see e.g. \cite{GuZai94,LarOv76+}). For the time independent 
component of this current we find 
\begin{equation} \label{CVC}
  I = \bar V/R_0 + I_{exc}(\omega,\tilde V,\bar V) + I_{Shapiro}(\omega,\tilde V,\bar V), 
\end{equation}
where $R_0$ is the Sharvin resistance of the junction, $I_{exc}$ is the additional
current due to multiple Andreev reflection and $I_{Shapiro}=$ $\sum_{k,n}I_{k,n}\,
\delta(2ne\bar V - k\hbar\omega)$ represents Shapiro peaks at the discrete constant
voltages $\bar V = (k/2n)\hbar\omega/e$. At low $T$ the latter includes subharmonics
($n>1$) because the current--phase relation strongly 
deviates from a standard sinusoidal form \cite{Kul70}. Below we shall 
focus our attention on the microwave field effect on the excess current
$I_{exc} = 2\int\,dE\,I_{\kappa=0}(E)$, where similarly to Eq. 
(\ref{f-kappa}) $I_{\kappa}$ is defined selfconsistently by
 \begin{eqnarray} \label{I-kappa}
  & & I_{\kappa}(E) = \sum_{k=-\infty}^{\infty} \, \sum_{\kappa'=-\infty}^{\infty} \, 
      m_{\kappa,\kappa'}(+,k,E) \\
  &\times& \left\{i_0        (E\!-\!e\bar V\!-\!k\hbar\omega)\delta_{\kappa',0} +  
                  I_{\kappa'}(E\!-\!e\bar V\!-\!k\hbar\omega)\right\}, \nonumber 
\end{eqnarray}
with $i_0(E) = (1/2eR_0)[{\cal A}(E)-1]\tanh(E/2k_B T)$. For $\tilde V \to 0$ the
Shapiro peaks disappear and Eq. (\ref{I-kappa}) is solved by the expression 
$I_{\kappa}(E) = \delta_{\kappa,0}\,\sum_{n=1}^{\infty} I_{n}(E)$ with 
\cite{OTBK,GuZai94,AvBard96}:
 \begin{eqnarray} 
  I_n(E) = \frac{1}{2eR_0}\prod_{l=0}^{n-1} {\cal A}(E\!-\!le\bar V) 
  \left[{\cal A}(E\!-\!ne\bar V)\!-\!1\right]\tanh[\frac{E\!-\!ne\bar V}{2k_BT}].
\end{eqnarray} 
\begin{figure}[t,h]
  \epsfbox{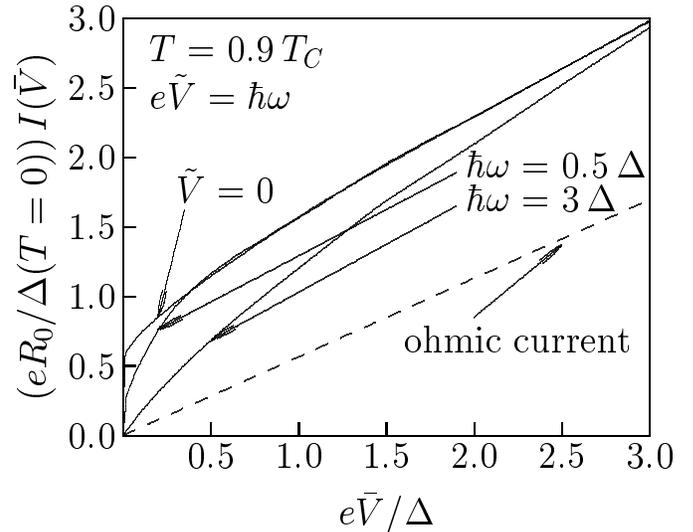}
\caption{CVC in the limit of constant voltage bias and in the presence of an 
additional microwave field for two different frequencies. Shapiro peaks
are not shown.}
\end{figure}
The I-V curves calculated numerically from Eqs. (\ref{CVC}) and (\ref{I-kappa}) 
for different $\omega$ are presented in Fig.~2 (Shapiro
peaks are not shown). For large voltages $e\bar V\gg\Delta,\hbar\omega$ 
one observes no influence of the microwave field on the excess current. 
With the aid of Eq. (\ref{I-kappa}) one can also demonstrate analytically 
that this result remains valid for $any$ microwave amplitude $\tilde V$.

The physical reason for this result is transparent. Absorbtion (emission)
of a photon in the weak link leads to a shift of the energy of 
quasiparticles by the value $\hbar\omega$. If only
one Andreev reflection takes place (i.e. for $e\bar V \gg \Delta$)
this shift is unique for all quasiparticles and drops out from the expression 
for the current. The same argument holds for multiphoton processes if 
the total energy shift is smaller than $e\bar V$. 

In contrast to the above limit, for smaller voltages $e\bar V < 2
\Delta$ we find a considerable suppression of the excess current 
$I_{exc}$. The effect becomes particularly pronounced at
large frequencies $\omega > 2\Delta$ (see Fig.~2). 
This result is a direct consequence of microwave induced 
$cooling$ and $heating$ of subgap quasiparticles with 
opposite momenta: the difference between ``$+$'' and ``$-$''  
distribution functions decreases (see Fig.~1) and so does
$I_{exc}$ (since the current $I$ is proportional to 
$\int dE[f_{+}(E-e\bar V/2)-f_{-}(E+e\bar V/2)]$).

To study this effect further let us calculate 
the low voltage conductance $G = 1/R_0 + 2\int\,dE\,G_{\kappa=0}(E)$ 
as a function of the microwave amplitude. For $\bar V\to 0$ 
one can derive from Eq. (\ref{I-kappa})
 \begin{eqnarray} \label{G-kappa}
  G_{\kappa}(E) &=& \sum_{k=-\infty}^{\infty}\,\sum_{\kappa'=-\infty}^{\infty} \, 
                    m_{\kappa,\kappa'}(s=+1,k,E) \\
  &\times& \left\{g_{0,\kappa'}(E\!-\!k\hbar\omega) +  
                  G_{\kappa'}  (E\!-\!k\hbar\omega)\right\}, \nonumber 
\end{eqnarray}
where $g_{0,\kappa}(E) = -e\,d[i_0(E)\,\delta_{\kappa,0} + 
I_{\kappa}(\bar V=0,E)]/dE$. 

For $\tilde V= 0$ the solution of Eq. (\ref{G-kappa}) reads $G_{\kappa}(E) = 
(1/2R_0){\cal A}(E)/[1-{\cal A}(E)] \, d[\tanh(E/2k_BT)]/dE \, \delta_{\kappa,0}$
and we reproduce the conductance peak $G-1/R_0 \sim (\ell_{in}/\xi_0)(1/R_0)
\tanh(\Delta/2k_BT)$ found in Ref. \cite{GuZai94}. The dependence $G(\tilde V)$ 
is shown in Fig.~3 for different microwave frequencies. We find a drastic 
decrease of $G(\tilde V)$ already for small microwave amplitudes $(e\tilde 
V/2\hbar\omega)^2 < 1$. For larger $\tilde V$ the effect becomes even more 
pronounced, multiple Andreev reflection turns out to be completely destroyed 
and the conductance is suppressed down to its normal state value $1/R_0$. 
\begin{figure}[t,h]
  \epsfbox{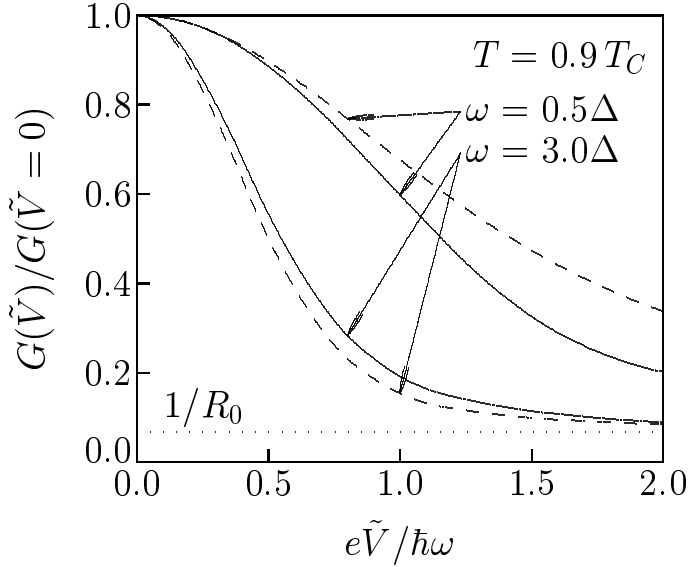}
\caption{Zero--bias--conductance $G(\tilde V)$ normalized to $G(\tilde V=0) = 
15/R_0$ versus amplitude $\tilde V$ of the applied microwave field. Dashed 
curves: $G(\tilde V)$ calculated without phase sensitive terms $m_{\kappa,
\kappa'}$, $\kappa\neq 0$.} 
\end{figure}


{\it Discussion.}
Our analysis of the microwave-induced effects in superconducting weak links
has led to a transparent physical picture which can be summarized as follows.
At low dc voltages subgap quasiparticles suffer multiple Andreev reflection 
increasing their energy by the value $e\bar V$ after each traverse across the 
weak link. Quasiparticle states with the momentum direction in (opposite to) 
that of the current become overpopulated (underpopulated) and the low voltage
conductance increases \cite{GuZai94}. The phase coherence is preserved
during the whole multiple Andreev reflection cycle. In the presence of an ac 
field subgap quasiparticles can also absorb and emit photons. These processes
destroy the phase coherence within a multiple Andreev reflection cycle and prevent
quasiparticles from further acceleration or, equivalently, lead to relative 
$cooling$ ($heating$) of ``$+$'' (``$-$'') subgap quasiparticles in comparison
to the case $\tilde V=0$ (see Fig.~1). As a result both the excess current
and the system conductance at low voltages can be strongly suppressed (Figs.~2,3). 
The suppression increases with the energy of absorbed and emitted photons $\hbar 
\omega$ and the amplitude of the microwave field $\tilde V$ in which case 
many-photon processes gain importance. This property makes it possible to use 
SNS structures as spectral microwave detectors. 
For larger voltages $e\bar V \gtrsim \Delta$ subgap quasiparticles
are accelerated and leave the weak link already after a few Andreev reflection
events. In this case the effect of an ac voltage becomes less pronounced. 
In the limit $e\bar V \gg \Delta$ only one Andreev reflection takes place 
and the I-V curve is not sensitive to an ac field.

In this paper we discussed the behavior of {\it voltage biased}
weak links in which case the low voltage ``foot'' structure of CVC is caused 
by multiple Andreev reflection. Although CVC of {\it current biased} weak links
shows a similar structure, in the latter case the ``foot'' is due to a
different physical reason -- the dc Josephson effect. In many experiments
the CVC of weak links is measured in the regime intermediate between the 
voltage and current biased limits. Therefore it may be quite difficult
to judge which of the above physical reasons could actually explain experimental 
results, in particular for short superconducting constrictions in which case
both the dc Josephson current and the low voltage excess current are of
the same order and have the same temperature dependence. 

The results obtained here suggest that a clear distinction between these two
mechanisms can be easily made in the presence of a microwave field. Indeed
in a microwave field (and at not very low $T$) $stimulation$ of the Josephson 
critical current takes place in superconducting microbridges \cite{AsLar76} 
and SNS junctions \cite{Zaikin83} whereas the low voltage excess current and 
the conductance become strongly $suppressed$. We believe that these opposite 
trends can be easily identified experimentally.

In conclusion, making use of the Keldysh technique we developed a microscopic
theory of nonequilibrium effects in superconducting weak links in the
presence of an external microwave field and demonstrated that these effects
may have a dramatic impact on charge transport in such structures. 

We acknowledge useful discussions with S.N. Artemenko, A.A. Golubov, N.Schopohl
and G. Sch\"on. The work was supported by SFB 195 of the Deutsche 
Forschungsgemeischaft.



\begin{references}
\bibitem{OTBK} M. Octavio, M. Tinkham, G.E. Blonder, and T.M. Klapwijk, Phys. 
Rev. {\bf B 27}, 6739 (1983). 
\bibitem{GuZai94} U. Gunsenheimer, and A.D. Zaikin, Phys. Rev. {\bf B 50}, 6317 
(1994); Physica {\bf B 203}, 280 (1994).
\bibitem{AvBard96} D. Averin, and A. Bardas, Phys. Rev. {\bf B 53}, R1705 (1996).
\bibitem{ZimKeck96} U. Zimmermann, and K. Keck, Z. Phys. {\bf B 101}, 555 (1996). 
\bibitem{LarOv76+} A.I. Larkin, and Yu.N. Ovchinnikov, Sov. Phys. JETP {\bf 41}, 960 
(1976); {\it ibid.} {\bf 46}, 155 (1977).
\bibitem{Schmid} see e.g. A. Schmid, in: {\it Nonequilibrium
superconductivity, phonons and Kapitza boundaries},
NATO Advanced Study Inst. Series B, Vol. 65, K.E. Gray (ed.), Plenum NY/London 
(1981).  
\bibitem{Zaikin83} A.D. Zaikin, 
Sov. Phys. JETP {\bf 57}, 910 (1983).
\bibitem{VolKlap92} This result is no longer valid in the
presence of inelastic relaxation and/or scattering on impurities in the
contact area. E.g. in the diffusive limit $\tilde f$ differs from zero 
and determines the charge transport across the weak link,
see e.g. A.F. Volkov, and T.M. Klapwijk, Phys. Lett. {\bf A 168}, 217
(1992), and Refs. therein. 
\bibitem{Kul70} I.O. Kulik, 
Sov. Phys. JETP {\bf 30}, 944 (1970); 
C. Ishii, Progr. Phys. {\bf 44}, 1525 (1970). 
\bibitem{AsLar76} L.G. Aslamazov, and A.I. Larkin, Zh. Eksp. Teor Fiz. 
{\bf 70}, 1340 (1976) [Sov. Phys. JETP {\bf 43}, 698 (1976)].
\end{references}
\end{document}